\documentclass{aa}
\usepackage{txfonts}
\usepackage{graphicx}
\usepackage{natbib}
\usepackage{amstext,cases}
\bibpunct{(}{)}{;}{a}{}{,}

\begin{document}

\title{On surface brightness fluctuations: probabilistic and statistical bases}
\subtitle{I: Stellar population and theoretical SBF}

\author{M. Cervi\~no\inst{1}, V. Luridiana\inst{1} \and L. Jamet\inst{1,2}}

\institute{Instituto de Astrof\'\i sica de Andaluc\'\i a (CSIC), Camino bajo
        de Hu\'etor 50, Apdo. 3004, Granada 18080, Spain
\and Instituto de Astronom\'\i a (UNAM), Apartado 70-264, M\'exico D.F., Mexico
}

\offprints{M. Cervi\~no, V. Luridiana, L. Jamet; \email{mcs@iaa.es, vale@iaa.es, ljamet@astroscu.unam.mx}}
\date{Received 20 March 2007; accepted 1 September 2008}

\abstract
{}
{This work aims to provide a theoretical formulation of Surface Brightness Fluctuations (SBF) in the framework of probabilistic synthesis models, in which there are no deterministic relations between the different stellar components of a population but only relations on average, and to distinguish between the different distributions involved in the SBF definition.}
{By applying the probabilistic theory of stellar population synthesis models, we estimate the shape (mean, variance, skewness, and kurtosis) of the distribution of fluctuations across resolution elements, and examine the implications for SBF determination, definition and application.}
{We propose three definitions of SBF: (i) stellar population SBF, which can be computed from synthesis models and provide an intrinsic metric of fit for stellar population studies; (ii) theoretical SBF, which include the stellar population SBF plus an additional term that takes into account the distribution of the number of stars per resolution element $\psi(N)$; theoretical SBF coincide with Tonry \& Schneider (1998) definition in the very particular case that $\psi(N)$ is assumed to be a Poisson distribution. However, the Poisson contribution to theoretical SBF is around 0.1\% of the contribution due to the stellar population SBF, so there is no justification to include any reference to Poisson statistics in the SBF definition; (iii) observational SBF, which are those obtained in observations that are distributed around the theoretical SBF. Finally, we show alternative ways to compute SBF and extend the application of stellar population SBF to defining a metric of fitting for standard stellar population studies.
}
{We demostrate that SBF are observational evidence of a probabilistic paradigm in population synthesis, where integrated luminosities have an intrinsic distributed nature, and they rule out the commonly assumed deterministic paradigm  of stellar population modeling.
}

\keywords{Galaxies: star clusters -- Galaxies: stellar content -- Methods: data analysis } 

\titlerunning{SBF: Probabilistic bases}

\maketitle

\section{Introduction}

Stellar population synthesis studies aim to decompose the integrated light, $L_{\mathrm{tot};\lambda}(N^*)$, of a sample of $N^*$ stars into a combination of particular stellar classes or populations, $n_i$. These studies can be developed by following two different methods. The first method \cite[][ as examples]{ST71,Fab72} completes a direct decomposition of the integrated light into principal components. The resulting solutions are {\it a posteriori} constrained by additional assumptions to obtain  the  description of the entire stellar population. In the second method  \cite[evolutionary stellar population synthesis, hereafter EPS; see][for an example]{Tin68}  the stellar populations are modeled following {\it a priori} rules related to stellar evolution and the relative contribution of each stellar population. Model results are compared with observations, and properties of the stellar populations (and their evolutionary status) of the targeted source are inferred. In both types of methods, degenerate solutions are present. The EPS method, however, is based on stellar evolution theory and provides both a tool for both studying the integrated light of galaxies and comparing its properties with stellar model predictions.

By construction, EPS needs to assume the existence of a probability distribution function that describes the mean value of the relative contribution of different populations to the integrated light, $\mu_1'(n_i)/N^*$. In addition, it assumes that each population can be characterized by a mean luminosity, $\mu_1'(l_{i;\lambda})$. EPS studies therefore refer to the {\it generic} emission of the integrated light of a set of systems rather than the {\it particular} emission of a given system. In a work inspired by \cite{GGS04}, \cite{CLCL06}, demonstrate that EPS provides a characterization of the probability distributions that describe {\it all} the possible integrated luminosities of an ensemble of stars with given evolutionary conditions\footnote{Throughout the paper, evolutionary conditions mean the age, metallicity, and star formation history of the stellar ensemble.}. In general, this characterization is expressed in terms of the {\it mean}, $\mu_1'(\ell_\lambda)$, of the distribution of the integrated luminosity of a system with a reference number of stars, $N^*_\mathrm{norm}$, that allows us to obtain the mean value of a distribution for a different number of stars, $N^*$, by simple transformations\footnote{In the following, we assume $N_\mathrm{norm}=1$ and we do not include the subindex $\lambda$ in the luminosities to simplify the notation. We also omit the subindex $\lambda$ in the following equations.}, $\mu_1'(L_{\mathrm{tot};\lambda};N^*) = \frac{N^*}{N^*_\mathrm{norm}} \, \mu_1'(\ell_\lambda)$. However, it is also possible to provide a more accurate characterization of such distributions using additional parameters of the distribution, such as the second raw moment, $\mu_2'(\ell_\lambda)$, the variance, $\kappa_2(\ell_\lambda)$, the skewness, $\gamma_1$, and the kurtosis, $\gamma_2$,  or even provide the distribution itself \citep{CLCL06}. This natural description of EPS in probabilistic terms, in which it is possible to establish the distribution of the integrated luminosity of a stellar ensemble, but not the precise position of this ensemble in the distribution, is not commonly considered in EPS literature, nor its application. A deterministic paradigm is usually applied:  $\mu_1'(n_i)/N^*$ is not a mean value but the actual fraction of stars of a given class present in the stellar system, regardless of the number of stars in the system. In the deterministic paradigm, given the total number of stars in the system and its evolutionary conditions, there is one and only one possible value of the integrated luminosity.

However, the probabilistic description of stellar populations,  although not recognized and studied in detail before, has been present in the literature since the late eighties.  \cite{TS88} presented the first description of the distribution of the integrated luminosities of stellar populations making use of the mean and the variance in the associated theoretical distributions, and defined theoretical Surface Brightness Fluctuations (SBF, $\bar{L}_{\mathrm{tot}}(N^*)$) as the variance of the integrated luminosity of a stellar population divided by its mean: $\bar{L}_{\mathrm{tot}}(N^*) = \kappa_2[{L}_{\mathrm{tot}}(N^*)]/\mu_1'[{L}_{\mathrm{tot}}(N^*)]$. Interestingly the dependence of $\kappa_2[{L}_{\mathrm{tot}}(N^*)]$ and $\mu_1'[{L}_{\mathrm{tot}}(N^*)]$ on the total number of stars cancels out numerically and $\bar{L}_{\mathrm{tot}}(N^*) = \bar{\ell}$ is independent of the number of the stars in the system. {\it Assuming a Poisson distribution} for the number of stars in each given stellar population, SBF can also be expressed as the ratio of the second to the first raw moments of the distribution of integrated luminosities, and in this case, it can be redefined to be the luminosity-weighted luminosity of the population.
 
 \cite{Buzz89} presented an independent probabilistic description  of EPS, also based on the variance of the corresponding distributions, and defined the effective number of stars, ${\cal{N}}(\ell)$, to be the ratio  $\mu_1'(\ell)^2/\kappa_2(\ell)$.  The  value of ${\cal{N}}[L_\mathrm{tot}(N^*)]$ for a system with $N^*$ stars scales directly with $\cal{N}(\ell)$ and it is an useful parameter for estimating sampling effects produced by the discreteness of real stellar populations. In his formulation, it was assumed that the relative number of stars in a given population fluctuates around the theoretical value  way described by a Poisson distribution \cite[but see][ and below]{CLCL06}. The connection between $\cal N$ and SBF was defined by \cite{Buzz93}, who related both quantities to the statistical entropy of stellar systems \cite[see also][ for a recent review]{Buzz07}.  

 As shown by \cite{TS88}, SBF is not only a theoretical concept, but a real observable that provides a relatively direct technique for determining extragalactic distances. The amplitude of the relative fluctuations in the observed flux with respect to a suitable mean is related directly to the theoretical SBF luminosity, and therefore to distance. In the observational domain, its principal asset is being able to disentangle the fluctuations that provide physical information about the system from the noise. Since they are related to evolutionary conditions, SBF provide information about stellar evolution and stellar population studies \cite[e.g.][]{CRBC03,GLLB07}.

Most work performed on theoretical SBF was related to its calibration as a distance indicator \cite[e.g.][]{TAL90}, although current calibrations are performed empirically. Further examples of SBF in stellar populations studies are \cite{W94,Con97,BCC98,BVA01}, and \cite{LCG00}. In the SBF definition, it is customary to assume that the number of stars in a given stellar population follows a Poisson distribution. Unfortunately, no complete study exists on theoretical SBF themselves and their implication in the use of EPS: the very observation of SBF implies that EPS models cannot be interpreted in a deterministic way, but as a description of a probability distribution that defines naturally an intrinsic {\it metric of fit} when physical properties are inferred from comparisons with model results \cite[see][ for more details]{CL07a,CL07b}.

On the other hand, the distributed nature described by EPS models was established clearly in theoretical studies of systems with a low number of stars, where the most probable number of stars in a given evolutionary stage $n_i/N^*$ differs significantly from the expected value i.e. the mean $\mu_1'(n_i)/N^*$. The characterization of integrated light in probabilistic terms began more recently. Partial solutions to the problems were presented by \cite{CLC00,BruTuc02,Gi02,LM00,CVGLMH02,Ceretal01,CVG03,CL04}, and \cite{FRDI07} among others, and a comprehensive solution to the problem was presented by \cite{CLCL06}, who introduced a probabilistic paradigm of EPS that applies to both under-sampled and well sampled systems.
 
In this paper, we establish a framework for theoretical SBF based on the probability distribution theory of stellar populations. The formalism allow us to describe the behavior of SBF for systems with a low number of stars per pixel, which was demonstrated by \cite{AT94} to be a limitation of the method.
Our main objective is to establish a robust definition of SBF. In particular, we show that the Poisson assumption is not essential to the SBF definition. We examine the implications of this robust definition and explore new ways to obtain observational SBF. Paraphrasing \citet{KP99}, ``once the definitions are laid out, the theory tends to fall into place, and understanding it is comparatively easy''.
 
The present paper is organized as follows. In Sect. \ref{sec:probabilityetc}, we summarize the theoretical probability distributions involved in SBF studies and provide an introduction to the probabilistic description of stellar populations. In Sect.~\ref{sec:theoreticalSBF}, we present a detailed
and quantitative description of SBF, establish the origin of SBF in probabilistic terms, and provide robust definitions of SBF according to the distributions involved. A discussion about the implications of the present result for SBF studies and additional SBF analysis methods are presented in Sect. \ref{sec:discussion}. Finally, our conclusions are given in Sect. \ref{sec:conclusion}.  In a companion paper (Cervi\~no et al. in preparation, hereafter Paper II), we will theoretically evaluate the observational error budget of the different ways to obtain SBF.

\section{The distributions involved in SBF studies}
\label{sec:probabilityetc}

Following \cite{TS88}, observational SBF are obtained from a statistical analysis of  fluctuations in the luminosity values of the pixels in the CCD image of a galaxy. First, a local mean is estimated at each point, and the fluctuations with respect to this mean are computed at each pixel. Second, the ratio between these fluctuations and the square root of the estimated local mean is computed at each point. Third,  the variance in the resulting distribution is the SBF of the image. The 
 \cite{TS88} method for obtaining SBF requires three distributions, although we will see in Sect. \ref{sec:SBFmodes} that SBF can be obtained using only two distributions. 
We note that {\it fluctuation} in {\it Surface Brightness Fluctuations} refers to the variance in {\it renormalized} fluctuations, i.e. the latter divided by the square root of the local mean. This renormalization is needed to ensure that all pixels are sampling distributions with a common mean and variance.  

Although the term {\it Surface Brightness Fluctuations} is understood to be a measure of the moments of the luminosity function from the analysis of the pixels in CCD images, we use a more general interpretation of SBF that relates any possible observation of renormalized fluctuations to the stellar population properties of the system.

The basic input is a set of observations of the same physical system. It can be the flux in a set of pixels from a galaxy (as proposed in the original method), a set of IFU obervations of a galaxy, a set of individual stars in a cluster or the integrated luminosities of different portions of a system. The only requirement is that all elements sample either the same luminosity distribution or a family of luminosities distributions whose parameters can be related to each other linearly (see below). In the following, we  define {\it resolution element} to be any element of this sample independently of whether we refer to, for example, a pixel, an IFU, an individual star.

The first distribution involved describes the possible integrated luminosities of a system for a given set of physical conditions, i.e. the luminosity distribution function of the population, or pLDF ($\varphi(L_{\mathrm{SP};N^*})$). This distribution is directly related to the stellar luminosity distribution function\footnote{This function provides the probability that a {\it single} star has a given luminosity, and it is only defined by the evolutionary conditions of a stellar sample.} (sLDF: $\varphi(\ell)$), by $N^*$ successive convolutions \cite[see][~for details]{CLCL06}. Implicitly, it is assumed that each resolution element is the result of a single random realization of $\varphi(L_{\mathrm{SP};N^*})$.

The second distribution involved is a reference distribution of {\it local} means (or, equivalently, expected values), needed to define fluctuations. This is the key point of observational SBF studies: when a system is observed, the number and class of stars in each resolution element are not a priori known, nor is the mean integrated luminosity corresponding to each resolution element. Hence, we need to {\it estimate} from the observation itself the expected flux that would correspond to each resolution element as opposed to the actual flux of the element. This estimation can be performed only from the statistical analysis of a set of resolution elements. The expected flux estimate itself depends on arbitrary observational/analysis choices, such as the number of resolution elements used in the estimation, and the choice of the region used to obtain the expected value, and might vary from one observation to another, or a data analysis to another (c.f. Sec. \ref{sec:SBFmodes}). 

Once the expected flux in each resolution element has been subtracted, the result is a pure distribution of fluctuations, that, by construction, has zero mean. However, the fluctuation of each resolution element depends on the number of stars in the element itself, which varies from element to element\footnote{The higher the number of stars, the larger the possible fluctuation, as we see below (cf. Eq. \ref{eq:varianceSSP}).}. So it is needed to renormalize the fluctuation by the square root of the expected flux. By this operation, the resolution elements sample a set of distributions that, by construction, share the same mean and variance. Observational SBF are just the {\it estimation} of the variance in this distribution of {\it renormalized} fluctuations.

In this work, we differentiate between {\it stellar population} SBF, {\it theoretical} SBF, and {\it observational} SBF. Stellar population SBF are described in terms of the basic physical process responsible for intrinsic luminosity fluctuations across resolution elements at fixed number of stars per resolution element. That is, the expected (mean) integrated flux in the resolution element is not obtained by observations but by a theoretical stellar population. 

Theoretical SBF are the SBF that can be theoretically predicted for a given system. They can be shown to contain a contribution from stellar population SBF plus a contribution from the distribution of the number of stars per resolution element. That is, the expected integrated flux in the resolution element is obtained from a theoretical model of the distribution of expected fluxes throughout the system. By their own definition, theoretical SBF measure a different physical quantity than stellar population SBF since they include an additional component. Finally, observational SBF are the observational counterpart of theoretical SBF.

As anticipated in the introduction, our model of the physical origins of SBF is based on the probabilistic description of stellar populations. Some basic concepts of this field that will be used in the paper are introduced in the following subsection, which is based on \citet{CLCL06} and may be skipped by those readers who are already familiar with that paper.

\subsection{The probabilistic description of stellar populations}\label{sec:probabilisticsynthesis}

We introduce a few concepts used in this work. The probabilistic description of SBF is based on the concept of stellar luminosity distribution function sLDF, which is denoted by $\varphi(\ell)$, where $\ell$ is the luminosity of individual stars. The sLDF provides the probability distribution of the luminosity of a randomly-selected {\it single} star. When referring to the sLDF, we will use $\mu_i(\ell)$ to indicate central moments (or moments about the mean) and $\mu'_i(\ell)$ to indicate raw moments (moments about zero). 

The explicit expressions for raw and central moments of the sLDF are:

\begin{eqnarray}
 \mu_n'(\ell)& =  & \int_0^\infty \ell^n \, \varphi(\ell) \, d\ell = \sum_{i}^{N_{\mathrm{ST}}} l_{i}^n \, p_{i}, \label{eq:2nd_momentp}\\
  \mu_n(\ell) & =  & \int_0^\infty (\ell - \mu_1')^n \,  \varphi(\ell) \, d\ell,  
 \label{eq:2nd_moment}
 \end{eqnarray}

\noindent where the mean $\mu_1'(\ell)$ is the first raw moment, and the variance $\mu_2(\ell)$ is the second central moment of the sLDF. In these equations, we have also expressed the integrals as a sum (as most synthesis codes do), where $l_{i}$ is the average luminosity of a given stellar type, $p_{i}$ the  probability density that a randomly-selected star belongs to stellar type $i$, and ${N_{\mathrm{ST}}}$ is the number of stellar types, such that $\sum_{i}^{N_{\mathrm{ST}}} p_{i} = 1$. 
We note that the stellar types should be defined in such a way that the luminosity can be assumed constant or to have a small variance for each given stellar type, i.e. $l_{i}$ is a truly representative value of the particular stellar type; the traditional classification of stars into evolutionary stages, such as the main sequence, and giant branch, is invalid in this context because the stellar luminosity spans a wide range at each of these stages.
 
We also use the cumulants $\kappa_n$ of different distributions.
The fundamental aspects of the use of cumulants in stellar population synthesis were described extensively in \cite{CLCL06} and will not be repeat in detail here. Interested readers may find an introduction to this topic in any advanced textbook of probability and statistics \cite[e.g.][]{KS77}.
We recall only the relations between the first four cumulants and moments:

\begin{eqnarray}
\kappa_1 & = & \mu_1',   \label{eq:kappa1}\\
\kappa_2 & = & \mu_2   = \mu_2' - \mu_1'^2,   \label{eq:kappa2}\\
\kappa_3 & = & \mu_3   =  \mu_3' - 3 \mu_1' \mu_2' + 2 \mu_1'^3, \label{eq:kappa3}\\
\kappa_4 & = & \mu_4 - 3 \mu_2^2  =  \mu_4' - 4 \mu_1' \mu_3' -3 \mu_2'^2 +12 \mu_1'^2 \mu_2'  - 6 \mu_1'^4,
\label{eq:kappa4}
\end{eqnarray}

\noindent and the definition of  the skewness, $\gamma_1$,  and the kurtosis, $\gamma_2$:

\begin{eqnarray}
\gamma_1 &= & \frac{\mu_3}{\mu_2^{3/2}} = \frac{\kappa_3}{\kappa_2^{3/2}}, \label{eq:gamma1} \\
\gamma_2 &= & \frac{\mu_4}{\mu_2^{2}} - 3= \frac{\kappa_4}{\kappa_2^{2}} . \label{eq:gamma2} 
\end{eqnarray}

When these symbols are expressed without a variate in parentheses or with the variate $\ell$, they refer to the sLDF. 

The same parameters can also be computed for the distribution of the integrated luminosity of a population, pLDF, which contains $N^*$ stars, $\varphi(L_{\mathrm{SP};N^*})$: in this case, we write the dependence explicitly, e.g. $\mu_1'(L_{\mathrm{SP};N^*})$. Integrated luminosities are indicated in upper case, e.g. $L_{\mathrm{SP};N^*}$ is the luminosity emitted by $N^*$ stars. Finally, we write the relation between the n-th order cumulant of the luminosity distribution of a cluster of $N^*$ stars, $\kappa_n(L_{\mathrm{SP};N^*})$, and the cumulants of the sLDF, $\kappa_n$: 

\begin{equation}
\kappa_n(L_{\mathrm{SP};N^*}) = N^*  \,  \times \, \kappa_n.
\label{eq:scale}
\end{equation}

We make extensive use of this equation, which can be found in advanced statistical books or, in the case of synthesis models of integrated populations, in \cite{CLCL06}. We note that the cumulants of pLDFs (but neither the raw moments nor central nth-moments with $n>3$) scale linearly with the number of stars. In this work, we use $\kappa_2$ instead  of $\mu_2$ to represent the variance of the distribution.

The skewness, $\Gamma_1$,  and kurtosis, $\Gamma_2$, of the pLDF with $N^*$ stars can be derived from Eqs.~\ref{eq:kappa1} $-$ \ref{eq:scale} and are given by:

\begin{eqnarray}
\Gamma_1(L_{\mathrm{SP};N^*}) &= & \frac{1}{\sqrt{N^*}} \gamma_{1}, \label{eq:Gamma1} \\
\Gamma_2(L_{\mathrm{SP};N^*}) &= & \frac{1}{N^*} \gamma_{2} . \label{eq:Gamma2}
\end{eqnarray}

\noindent Since for Gaussian distributions $\Gamma_1=\Gamma_2=0$, Eqs.~\ref{eq:Gamma1} and \ref{eq:Gamma2} indicate that the higher the number of stars $N^*$, the more Gaussian-like the distribution becomes.

Finally, we use $N^*$ to denote the number of stars per resolution element when this is a constant  (stellar population case); $N_{\Sigma_\mu}$ to represent the {\it mean} number of stars per resolution element in the area $\Sigma$ used to estimate the mean luminosity in a resolution element; $N_{\mathrm{pois}}$ to represent $N_{\Sigma_\mu}$  when the number of stars per resolution element is found to be represented by a Poisson distribution in the area considered; and $N$ to refer to the number of stars per resolution element when this is a variable quantity. 

\section{Theoretical definitions of SBF}\label{sec:theoreticalSBF}

In Sect.~\ref{sec:probabilityetc} we emphasize the necessity of describing the theoretical SBF definition of \cite{TS88} arising from two factors. The first is the intrinsic dispersion in the total luminosity of a population of a given number of stars $N$; the second is the distribution of number of stars in our resolution elements, which is needed to define a reference value to define the fluctuation.

The first of these factors is addressed in the theoretical framework of probabilistic synthesis models:  a particular value of the integrated light  is just a realization of all possible random combinations of a given number of individual stars with the same evolutionary conditions. Different realizations consist of different stellar mixes\footnote{This is a simplification of the original theory, in which IMF sampling is but one of several causes of dispersion; it is, however, sufficient to develop our argument. See \citet{CLCL06} for more details.}, and display different integrated luminosities. Even in the case of a strictly constant total number of stars in all resolution elements considered, fluctuations in the stellar mix  {\it always} introduce a dispersion in the integrated luminosity values. In the cases of an observation of an extended system with homogeneous evolutionary conditions,  each resolution element provides the integrated luminosity of different realizations of the same (theoretical) stellar population. In our framework, we define {\it stellar population SBF} to be those arising from this effect. The expression for population SBF is derived in the following.

\subsection{Population SBF}\label{sec:populationSBF}

We assume that all resolution elements have exactly the same number of stars, $N^*$. In such a case, the luminosity of each element is a variable randomly-selected from the pLDF. The first step in computing SBF consists of the characterization of the pLDF, $\varphi(L_{\mathrm{SP};N^*})$, the parent distribution of the $L_{\mathrm{SP};N^*}$, which is the luminosity of a resolution element with $N^*$ stars.  Equation \ref{eq:scale} indicates that the mean luminosity and variance of the pLDF is related with the mean, $\mu_{1}'$, and variance, $\kappa_{2}$, of the sLDF by:

\begin{eqnarray}
\mu_{1}'(L_{\mathrm{SP};N^*}) &=& N^* \, \mu_{1}', \\
\kappa_{2}(L_{\mathrm{SP};N^*}) &=& N^* \, \kappa_{2}. 
\label{eq:varianceSSP}
\end{eqnarray}

As we have emphasized before, the higher the value of $N^*$, the larger $\kappa_{2}(L_{\mathrm{SP};N^*})$, or equivalently, the larger the amplitude of the possible fluctuations. Additionally, the theoretical scatter, $\sigma(L_{\mathrm{SP};N^*}) = \sqrt{\kappa_{2}(L_{\mathrm{SP};N^*})}$ scales with $\sqrt{N^*}$ and
the mean integrated luminosity scales with $N^*$. Hence, it is natural to renormalize the resulting fluctuation by $\sqrt{\mu_{1}'(L_{\mathrm{SP};N^*})}$ to eliminate numerically this dependence on $N^*$. The distribution  of the renormalized luminosity fluctuations, which is indicated by $\varphi(L^\mathrm{fluc}_{\mathrm{SP};N^*})$, is the distribution of the variable:

\begin{eqnarray}
L^\mathrm{fluc}_{\mathrm{SP};N^*} &=& \frac{L_{\mathrm{SP};N^*} - \mu_{1}'(L_{\mathrm{SP};N^*})}{\sqrt{\mu_{1}'(L_{\mathrm{SP};N^*})}} 
= \frac{L_{\mathrm{SP};N^*} -N^* \mu_{1}'}{\sqrt{N^* \, \mu_{1}'}},
\label{eq:phiSBFteo}
\end{eqnarray}

\noindent i.e. it can be obtained from $\varphi(L_{\mathrm{SP};N^*})$ by means of a linear transformation. Using Eq.~\ref{eq:kappa2}, the population SBF associated with resolution elements containing $N^*$ stars, i.e. the variance of $\varphi(L^\mathrm{fluc}_{\mathrm{SP};N^*})$, is:

\begin{eqnarray}
\bar{L}_{\mathrm{SP};N^*} &=& \kappa_{2}(L^\mathrm{fluc}_{\mathrm{SP};N^*}) = \frac{\kappa_{2}(L_{\mathrm{SP};N*})}{N^* \, \mu_{1}'} = \frac{N^* \, \kappa_{2}}{N^* \, \mu_{1}'} = \frac{\kappa_{2}}{\mu_{1}'}.
\label{eq:SBFN*}
\end{eqnarray}

\noindent where $\bar{L}_{\mathrm{SP};N^*}$ is numerically independent of the number of stars per resolution element  (assuming that all resolution elements in this region have exactly the same number of stars). Hence, we can omit the subindex $N^*$ and assume that the general expression for population SBF is:

\begin{equation}
\bar{L}_{\mathrm{SP}} \equiv \frac{\kappa_{2}}{\mu_{1}'}.
\label{eq:SBFteo}
\end{equation}

However, although $\bar{L}_{\mathrm{SP}}$ is independent of $N^*$, this does not mean that the shape of the  distribution of renormalized fluctuations does not depend on $N^*$. The shape of this distribution can be inferred by computing its skewness and kurtosis. The 3rd and 4th cumulants of $\varphi(L^\mathrm{fluc}_{\mathrm{SP};N^*})$ are:

\begin{eqnarray}
\kappa_{3}(L^\mathrm{fluc}_{\mathrm{SP};N^*}) &=&\frac{\kappa_{3}(L_{\mathrm{SP};N*})}{\sqrt{(N^*\, \mu_{1}')^3}} = \frac{N^*\, \kappa_{3}}{\sqrt{(N^* \, \mu_{1}')^3}} = \frac{\kappa_{3}}{\sqrt{N^* \, \mu_{1}'^3}},
\label{eq:k3SBFteo} \\
\kappa_{4}(L^\mathrm{fluc}_{\mathrm{SP};N^*}) &=&\frac{\kappa_{4}(L_{\mathrm{SP};N*})}{(N^*\, \mu_{1}')^2} = \frac{N^* \, \kappa_{4}}{(N^*\, \mu_{1}')^2} = \frac{\kappa_{4}}{N^* \, \mu_{1}'^2}, 
\label{eq:k4SBFteo}
\end{eqnarray}

\noindent and the skewness and kurtosis are therefore :

\begin{eqnarray}
\Gamma_{1}(L^\mathrm{fluc}_{\mathrm{SP};N^*}) &=& \frac{\kappa_{3}}{\sqrt{N^* \, \mu_{1}'^3}} \, \left(
\frac{\kappa_{2}}{\mu_{1}'}\right)^{-3/2} = \frac{1}{\sqrt{N^*}} \, \gamma_{1},
\label{eq:g1SBFteo}\\
\Gamma_{2}(L^\mathrm{fluc}_{\mathrm{SP};N^*}) &=& \frac{\kappa_{4}}{N^* \, \mu_{1}'^2} \, \left(
\frac{\kappa_{2}}{\mu_{1}'}\right)^{-2} = \frac{1}{N^*} \, \gamma_{2}.
\label{eq:g2SBFteo}
\end{eqnarray}

We note that Eqs.~\ref{eq:g1SBFteo} and \ref{eq:g2SBFteo} provide the same values as Eqs.~\ref{eq:Gamma1} and \ref{eq:Gamma2}, so the shape in terms of the skewness and kurtosis of $\varphi(L^\mathrm{fluc}_{\mathrm{SP};N^*})$ is identical to that of the pLDF of a population of $N^*$ stars, $\varphi_{\mathrm{L_{\mathrm{tot}}}}(L_{\mathrm{SP};N^*})$. Equations.~\ref{eq:g1SBFteo} and \ref{eq:g2SBFteo} also indicate that the shape of the distribution of renormalized fluctuations depends on the number of stars per resolution element, although all distributions share the same mean and variance. As an example, \cite{CLCL06} demonstrated that $N^*$ must have values between $10^5$ (in visible bands) and $10^7$ (in infrared bands) for Gaussian distributions of both the pLDF and renormalized fluctuations, to be achievable. That is, a given system would have a Gaussian distribution of renormalized fluctuations at one wavelength and a non-Gaussian distribution at other wavelengths. For a more detailed study of the number of stars needed to reach a Gausssian regime, we refer to the more detailed treatment presented in \cite{CLCL06}.

\subsection{Theoretical SBF: effect of a variable number of stars per resolution element in the estimation of local means}\label{sec:localeffect}

The population $\bar{L}_{\mathrm{SP}}$ (Eq. \ref{eq:SBFN*}) is the theoretical SBF in the case in which all resolution elements have the same number of stars. In the general case, however, the number of stars per resolution element will differ for different elements and each element will sample a different pLDF. Hence, the local mean estimated from resolution elements in a region $\Sigma_\mu$ depends on the distribution of the number of stars in the resolution elements included in the region.

We now derive the theoretical expression of SBF for the general case of estimation of a mean for resolution elements with a variable number of stars. In doing so, we must of course mention the set of resolution elements for which the local mean is computed, since this defines the distribution of the number of stars. We therefore indicate this region with $\Sigma_\mu$, and introduce an additional distribution that represents the distribution of the variable number of stars per resolution element, $N$, in this region, $\psi_{\Sigma_\mu}(N)$. This distribution indicates the probability of having a resolution element with $N$ stars, or, in frequency terms, the relative number of resolution elements with $N$ stars when the number of resolution elements is infinity.  We assume that $\int \psi_{\Sigma_\mu}(N)\,dN = 1$. In the remainder of Sect. \ref{sec:localeffect}, we determine how the mean and variance of the fluctuation image depend {\it locally} on $\psi_{\Sigma_\mu}(N)$. These results are used in Sect. \ref{sec:globaleffect} to derive a general expression for SBF.

The distribution of the possible luminosity values for the set of resolution elements considered, $\phi(L_{\Sigma_\mu})$, is the result of the sum of all possible $\varphi(L_{\mathrm{SP};N})$ distributions, weighted by the distribution of the number of stars $N$ in the set of resolution elements defined by this area:

\begin{equation}
\phi(L_{\Sigma_\mu}) = \int_{\Sigma_\mu} \,\psi_{\Sigma_\mu}(N) \, \varphi(L_{\mathrm{SP};N}) \, dN. 
\label{eq:PDL-local}
\end{equation}

After  simple but cumbersome algebraic operations, the following is obtained:

\begin{eqnarray}
\mu_{1}'(L_{\Sigma_\mu}) &=& \mu_{1}'(N)\, \mu_{1}' (\ell)\label{eqs:k1local}, \\
\kappa_{2}(L_{\Sigma_\mu}) &=& \mu_{1}'(N) \, \kappa_{2}(\ell)  + \kappa_{2}(N)\, \mu_{1}'^2(\ell)  \label{eqs:k2local},  \\
\kappa_{3}(L_{\Sigma_\mu}) &=& \mu_{1}'(N) \, \kappa_{3}(\ell)  + 3\kappa_{2}(N)\,\mu_{1}' (\ell)\, \kappa_{2}(\ell) + \kappa_{3}(N)\,\mu_{1}'^3(\ell)  \label{eqs:k3local},  \\
\kappa_{4}(L_{\Sigma_\mu}) &=& \mu_{1}'(N) \, \kappa_{4}(\ell) + \kappa_{2}(N)\,[3 \kappa_{2}^2(\ell) + 4\mu_{1}'(\ell)\,\kappa_{3}(\ell)] + \nonumber \\
&&6\kappa_{3}(N)\,\mu_{1}'^2(\ell)\,\kappa_{2}((\ell) + \kappa_{4}(N)\,\mu_{1}'^4(\ell),
\label{eqs:k4local}
\end{eqnarray}

\noindent where $\kappa_{i}(N)$ are the cumulants of the distribution $\psi_{\Sigma_\mu}(N)$, and $\kappa_{i}(\ell)$ are the cumulants of the sLDF. 

The transformation 

\begin{equation}
L^\mathrm{fluc}_{\Sigma_\mu} = \frac{L_{\Sigma_\mu} - \mu_{1}'(L_{\Sigma_\mu})}{\sqrt{\mu_{1}'(L_{\Sigma_\mu})}}
\end{equation}

\noindent described previously now provides the distribution $\phi(L_{\Sigma_\mu}^\mathrm{fluc})$ with $\mu_1'(L_{\Sigma_\mu}^\mathrm{fluc})=0$ and $\kappa_2(L_{\Sigma_\mu}^\mathrm{fluc})$ (i.e., the `local' SBF) given by:

\begin{equation}
\bar{L}_{\Sigma_\mu} = \kappa_{2}(L_{\Sigma_\mu}^\mathrm{fluc})=\frac{\kappa_{2}(L_{\Sigma_\mu})}{\mu_{1}'(L_{\Sigma_\mu})} = \bar{L}_{\mathrm{SP}} + \mu_{1}'(\ell)\, \frac{\kappa_{2}(N)}{\mu_{1}'(N)},
\label{eq:SBFgeneral}
\end{equation}

\noindent and a skewness and kurtosis given by

\begin{eqnarray}
\Gamma_{1}(L_{\Sigma_\mu}^\mathrm{fluc}) & = &\frac{\kappa_{3}(L_{\Sigma_\mu})}{\sqrt{\kappa_{2}(L_{\Sigma_\mu})^3}}, \\  
\Gamma_{2}(L_{\Sigma_\mu}^\mathrm{fluc}) & = &\frac{\kappa_{4}(L_{\Sigma_\mu})}{\kappa_{2}(L_{\Sigma_\mu})^2}.   
\end{eqnarray}

In general, the contribution from the variation in the number of stars per resolution element in the region $\Sigma_\mu$ in which the reference mean is deriveed must be modeled and evaluated for each system if one wants to compare the SBF with stellar population SBF. However, in the case of relatively close galaxies with smooth luminosity profiles, it can be assumed that the luminosity profile in any region $\Sigma_\mu$ is almost flat and that $\psi_{\Sigma_\mu}(N)$ follows a Poisson-like distribution. We  consider this specific case in the following section.

\subsubsection{Case of Poisson distribution in the number of stars}

There are two cases in which the above equations provide a  simple relation to the moments of the sLDF: (a) $\psi_{\Sigma_\mu}(N)$ is a Dirac delta function centered on $N^*$; (b) $\psi_{\Sigma_\mu}(N)$ is a Poisson distribution with mean value $N_{\mathrm{pois}}$. The first of these cases corresponds to the $N=N^* = \mathrm{constant}$ case discussed above. In the Poisson case, it is known that $\mu_{1}'(N)=\kappa_{2}(N)=  \kappa_{3}(N) = \kappa_{4}(N)=N_\mathrm{pois}$. Introducing these relations in Eqs.~\ref{eqs:k1local} to \ref{eqs:k4local}, we obtain:

\begin{eqnarray}
\mu_{1}'(L_{\Sigma_\mu;N_{\mathrm{pois}}}) &=& N_\mathrm{pois} \, \mu_{1}'  \label{eqs:k1localPoi}, \\
\kappa_{2}(L_{\Sigma_\mu;N_{\mathrm{pois}}}) &=& N_\mathrm{pois} (\kappa_{2} + \mu_{1}'^2) = N_\mathrm{pois} \, \mu_{2}' \label{eqs:k2localPoi},\\
\kappa_{3}(L_{\Sigma_\mu;N_{\mathrm{pois}}}) &=& N_\mathrm{pois} \,\mu_{3}' \label{eqs:k3localPoi}, \\
\kappa_{4}(L_{\Sigma_\mu;N_{\mathrm{pois}}}) &=& N_\mathrm{pois} \, \mu_{4}'.
\label{eqs:k4localPoi}
\end{eqnarray}

These equations characterize the distribution of luminosities in the case in which the distribution of the number of stars per resolution element follows a Poisson distribution with mean $N_\mathrm{pois}$. We note that $\kappa_{2}(L_{\Sigma_\mu;N_{\mathrm{pois}}})$ contains a component equal to the variance of a pLDF with $N_{\mathrm{pois}}$ stars, plus a component equal to the variance in luminosity of a Poisson distribution of identical stars with individual luminosity given by $\mu_{1}'$. We can characterize  the distribution of renormalized fluctuations $\phi(L_{\Sigma_\mu;N_{\mathrm{pois}}}^\mathrm{fluc})$ by:

\begin{eqnarray}
\bar{L}_{\Sigma_\mu;N_{\mathrm{pois}}} &=& \bar{L}_{\mathrm{SP}} + \mu_{1}' = \frac{\mu_{2}'}{\mu_{1}'}, \label{eq:SBFPoi} \\
\Gamma_{1}(L^\mathrm{fluc}_{\Sigma_\mu;N_{\mathrm{pois}}}) &=& \frac{1}{\sqrt{N_\mathrm{pois}}} \frac{\mu_{3}'}{\mu_{2}'^{3/2}},
\label{eq:g1SBFpoiteo}\\
\Gamma_{2}(L^\mathrm{fluc}_{\Sigma_\mu;N_{\mathrm{pois}}}) &=& \frac{1}{N_\mathrm{pois}} \frac{\mu_{4}'}{\mu_{2}'^2}.
\label{eq:g2SBFpoiteo}
\end{eqnarray}

\noindent Equation \ref{eq:SBFPoi} is a rigorous proof of the assertion of \cite{TS88} that the SBF can be calculated to be the ratio of the first to the second raw moment of the stellar luminosity function for a Poissonian approximation to the distribution of the number of stars per resolution element used to compute the mean flux of each resolution element. However, we note that a Poissonian distribution of the number of stars per resolution element must not be taken for granted in all situations.

\subsection{Theoretical SBF: global effect of a variable number of stars in the resolution elements}\label{sec:globaleffect}

In the previous case, we obtained the distribution of possible renormalized fluctuations in a `local' region that includes a given resolution element and where the mean had been obtained from the analysis of an additional set of resolution elements defining a region $\Sigma_\mu$. In this section, we derive the expression for theoretical SBF by assuming that the local luminosity mean values are obtained as above and integrating the local SBF over the entire system. Generally speaking, the distribution of stars per resolution element in each $\Sigma_\mu$ throughout the galaxy will vary and there will be a different set of Eqs. \ref{eqs:k1local} to \ref{eqs:k4local} for each resolution element. That is, we have a $\psi_{\Sigma_\mu}(N)$ distribution defined for each resolution element. Since the problem is difficult to solve in its generality, to simplify we assume that the four cumulants of $\psi_{\Sigma_\mu}(N)$ can be expressed as linear functions of the mean number of stars per resolution element, $N_{\Sigma_\mu}$, inside each $\Sigma_\mu$ (which may vary from element to element). Implicit in this hypothesis is that the cumulants that depend on the stellar population ($\mu_1(\ell)$, $\kappa_2(\ell)$, etc.) are the same for all resolution elements, i.e. the stellar population is homogeneous across the region in which the final SBF are computed, or  the sLDF is universal, or  any variation in the stellar population properties occurs smoothly. We note that this hypothesis is implicit in any SBF analysis and is not a particular assumption of this work. 

With this hypothesis, we define $\Sigma_{\mathrm{SBF}}$ to be the region in which final SBF are evaluated; it can be the entire or a region of the galaxy. The distribution of renormalized fluctuations in which we are interested, $\Phi_{\Sigma_\mathrm{SBF}}(L^\mathrm{fluc})$, is the result of the sum of all possible $\phi(L^{\mathrm{fluc}}_{\Sigma_\mu})$ distributions weighted by the distribution of $N_{\Sigma_\mu}$, given by $\Psi_{\mathrm{SBF}}(N_{\Sigma_\mu})$, which is related to the  luminosity profile of the galaxy region over which the SBF is computed. 

The distribution of renormalized fluctuations can be obtained as in the previous case:

\begin{equation}
\Phi_{\Sigma_\mathrm{SBF}}(L^\mathrm{fluc}) = \int_{\Sigma_{\mathrm{SBF}}} \,\Psi_{\mathrm{SBF}}(N_{\Sigma_\mu})\, \phi(L^{\mathrm{fluc}}_{\Sigma_\mu}) \, dN_{\Sigma_\mu}. 
\label{eq:SBF-profile}
\end{equation}

We note that this equation is formally equal to Eq.~\ref{eq:PDL-local}, but in place of the local distribution of the number of stars from Eq.~\ref{eq:PDL-local}, we have the global distribution of the {\it mean} number of stars. Furthermore, in Eq.~\ref{eq:PDL-local} we had the pLDF, whereas in Eq.~\ref{eq:SBF-profile} we have the distribution of renormalized fluctuations $\phi(L^{\mathrm{fluc}}_{\Sigma_\mu})$. This is necessary because of the limitations intrinsic to the observational method: since the local mean and the distribution of stars per resolution element are unknown, the local mean must be obtained from a set of resolution elements, then used to obtain the fluctuation image (which depends on the definition of the local area $\Sigma_\mu$), with which, finally, the SBF are obtained.

\paragraph{Case of constant number of stars in all resolution elements.}
Let us obtain the cumulants of this distribution in the theoretical case of a fixed number of stars per resolution element  $\phi(L^{\mathrm{fluc}}_{\Sigma_\mu}) = \varphi_{\mathrm{L_{\mathrm{tot}}}}(L_{\mathrm{SP};N^*}^\mathrm{fluc})$. We must perform the same algebraic operations as in the previous case, but by using instead the fluctuation distribution of the pLDF. In the case of $\kappa_{2}[\Phi_{\Sigma_\mathrm{SBF}}(L^\mathrm{fluc}_{\mathrm{SP};N^*})]$, we find:

\begin{equation}
\bar{L}=\kappa_{2}[\Phi_{\Sigma}(L^\mathrm{fluc}_{\mathrm{SP};N^*})] = \frac{\kappa_{2}}{\mu_{1}'} = \bar{L}_{\mathrm{SP}},\\
\end{equation}

\noindent and the skewness and kurtosis of $\Phi(L^\mathrm{fluc}_{\mathrm{SP};N^*})$ are:

\begin{eqnarray}
\Gamma_{1}[\Phi_{\Sigma}(L^\mathrm{fluc}_{\mathrm{SP};N^*})] &=&  \gamma_{1} \left< \frac{1}{\sqrt{N^*}} \right>_{\mathrm{SBF}} \label{eq:Gamma1SBFtot},\\
\Gamma_{2}[\Phi_{\Sigma}(L^\mathrm{fluc}_{\mathrm{SP};N^*})] &=&  \gamma_{2} \left< \frac{1}{N^*} \right>_{\mathrm{SBF}}, \label{eq:Gamma2SBFtot}
\end{eqnarray}

\noindent where the bracket notation $<x>_{\mathrm{SBF}}$ denotes the expected value of $x$ over the region in which the SBF is computed.

\paragraph{Case of Poisson-distributed number of stars.}
If we assume that the number of stars per resolution element in the regions we use to obtain the mean values follows a Poisson distribution, $\phi(L^{\mathrm{fluc}}_{\Sigma_\mu}) = \phi(L^{\mathrm{fluc}}_{\Sigma_\mu;N_{\mathrm{pois}}})$, the following relations are obtained:

\begin{eqnarray}
\bar{L} = \kappa_{2}[\Phi_{\Sigma}(L^\mathrm{fluc}_{\mathrm{pois};N_{\mathrm{pois}}})] &=& \frac{\mu_{2}'}{\mu_{1}'} = \bar{L}_{\Sigma_\mu;\mathrm{pois}}\label{eq:SBFpoi}, \\
\Gamma_{1}[\Phi_{\Sigma}(L^\mathrm{fluc}_{\mathrm{pois};N_{\mathrm{pois}}})] &=&  \frac{\mu_{3}'}{\mu_{2}'^{3/2}} \left< \frac{1}{\sqrt{N_{\mathrm{pois}}}} \right>_{\mathrm{SBF}}, \\
\Gamma_{2}[\Phi_{\Sigma}(L^\mathrm{fluc}_{\mathrm{pois};N_{\mathrm{pois}}})] &=&  \frac{\mu_{4}'}{\mu_{2}'^{2}} \left< \frac{1}{N_{\mathrm{pois}}} \right>_{\mathrm{SBF}}.
\end{eqnarray}

In both cases, the shape of $\Phi_{\Sigma}(L^\mathrm{fluc})$ depends on $\langle 1/\sqrt{N} \rangle$ and $\langle 1/N \rangle$. We note that these values are in general different from $1/\sqrt{\langle N \rangle}$ and $1/{\langle N \rangle}$. In particular, it can be shown \cite[][ exercise 9.13]{KS77} that the product $\langle N \rangle \times \langle 1/N \rangle  \geq 1$ if $N > 0$, as in our case. The equality only holds if the distribution of $N$ is a Dirac delta function. Hence, in general, the use of $1/\langle N \rangle$ provides only lower limits to the estimation of the shape of  $\Phi_{\Sigma}(L^\mathrm{fluc})$.

\paragraph{Implications in the general case.}
An analogous reasoning, but leading to more complicated expressions, applies to the general case. 
There are two relevant aspects here: (1) The global SBF coincides with the local SBF. Since the latter contains the contribution of the stellar population SBF plus the contribution of the variation in the total number of stars in the region used to obtain the reference mean value, the final SBF also includes these contributions. (2) In general, the distribution of renormalized fluctuations, $\Phi_{\Sigma}(L^\mathrm{fluc})$, does not follow a Gaussian distribution and its shape will depend on the value of $<1/\sqrt{N}>$ and $<1/N>$ and the raw moments or cumulants of the sLDF. Only for a sufficiently high number of stars {\it per resolution element} will this distribution be Gaussian. A quantitative evaluation of how high this number can be, is found in \cite{CLCL06}. As reference values, $\Gamma_1$ and $\Gamma_2$ must be lower than 0.05 to achieve a Gaussian-like symmetric distribution (at a $3 \sigma$ level), and $\Gamma_1$ and $\Gamma_2$ must be lower than 0.5 to avoid a bimodal distribution of renormalized fluctuations. As an example, a Gaussian distribution of renormalized fluctuations is obtained for systems with more than $10^7$ stars per resolution element (e.g. pixel) for old stellar clusters in infrared bands.

We now explore the shape of the distribution of renormalized fluctuations for the  $z$ band used in HST-ACS surveys \citep{Meietal07}. 
Assuming that the typical values of skewness and kurtosis of the sLDF for the $z$ band of an old stellar population are simillar to those of the I band, we use the results of \cite{CLCL06}  where  $\gamma_{1}(I) \approx 10^2$ and $\gamma_{2}(I) \approx 10^3$.

For the Virgo cluster observed by the ACS/WFC, the projected distance in a pixel is $\approx 4$ pc 
assuming a distance of $D\approx16.5$ Mpc and a pixel scale $\theta \approx 0.05\arcsec/\mathrm{pix}$. Assuming an average projected density of stars of $10^3$ stars pc$^{-2}$ \citep{TL06}, the number of stars per pixel is $\langle N \rangle \approx 1.6 \times 10^4$. This number yields values of $\Gamma_{1}[\Phi_{\Sigma}(I)] \approx  0.8$ and  $\Gamma_{2}[\Phi_{\Sigma}(I)] \approx  0.06$. That is, the distribution of renormalized fluctuations is asymmetric and possibly multimodal.  It is also possible to use a larger resolution element for SBF computation corresponding more closely to the ACS/WFC $z$ band PSF, which can be approximated by areas of 3x3 pixels. In this case, both  $\Gamma_{1}[\Phi_{\Sigma}]$ and  $\Gamma_{2}[\Phi_{\Sigma}]$ decrease by factors of 3 and 9 respectively, which yields $\Gamma_{1}[\Phi_{\Sigma}(I)] \approx 0.26$, still a factor of 5 higher than that required to ensure gaussianity.

Considering M32 and \cite{TS88}, we assume a distance of 0.9 Mpc and a pixel scale of $\theta \approx 0.415\arcsec/\mathrm{pix}$; the projected distance in a pixel is then 1.8 pc, which yields $\langle N \rangle \approx 3.5 \times 10^3$. In the V band, with $\gamma_{1}(V) \approx 50$ and $\gamma_{2}(V) \approx 400$ \citep{CLCL06}, it is found $\Gamma_{1}[\Phi_{\Sigma}(V)] \approx  0.8$ and  $\Gamma_{2}[\Phi_{\Sigma}(V)] \approx  0.1$. This distribution is again non-Gaussian, in contrast to the assumption made by \cite{TS88} based on an estimated frequency of 20 giant stars per pixel. We note that giants are a {\it broad} stellar type, which cover a broad range in luminosity;  taken at face value this number, does not therefore provide sufficient information about the shape of the distribution of renormalized fluctuations.

\section{Discussion}
\label{sec:discussion}

Once the distribution involved in SBF has been obtained, it is possible to propose a robust definition of theoretical SBF, establish guidelines for comparisons between theoretical and observational  SBF, propose  additional methods to obtain SBF and extend its range of applications. In this section, we first discuss some issues related to SBF definitions found in the literature (in particular the unnecessary mention of Poisson distributions). Secondly, we propose additional methods to obtain SBF and discuss the advantages and differences. Finally, we show some additional applications of the method.

\subsection{Poisson or not Poisson?}

The most common explanation of the origin of SBF found in the literature is Poisson fluctuations in the number of stars (and in the number of stars per stellar type in a given population) across different resolution elements. However, we have shown that in this type of definition two different concepts are entangled: 

\begin{enumerate}
\item  The first concept involved is the population SBF, which appear naturally as a manifestation of the distribution of the possible luminosity of any ensemble {\it with a fixed number of stars} and physical conditions. 

\item The second concept involved is the distribution of the {\it total} number of stars in different resolution elements that are needed to obtain a reference value for the definition of fluctuations. 
We have shown that it can be understood as the possible fluctuation when a population consists only of a particular type of stars, all of the same luminosity $\ell$. That is, the stellar population itself does not produce any fluctuation at all, and the only component of the SBF arises as a consequence of variations in the total number of stars. In this case, the SBF would be $\ell$ if the total number of stars per resolution elements follow a Poisson distribution \citep{TS88}, although a more general, exact, and simple formulation can be obtained without the use of Poisson approximation. 

\end{enumerate}

In our work, we have assumed that (1) the number of stars per resolution element is fixed (and produce the stellar population SBF) although (2) it is impossible to determine the number of stars  per resolution element and we must consider instead  an average value. 
When \cite{TS88} defined theoretical SBF on a Poisson base, they used these two different contributions implicitly: that of the stellar population SBF and that due to the local variation in the mean number of stars per resolution element (where the Poisson distribution applies), but without a clear distinction between both components. We emphasize again that \cite{TS88} make use of a probabilistic description of EPS since they assume that the number of stars of a given stellar type does not have a fixed value, but fluctuates about a mean. Under a deterministic paradigm of EPS, where for given evolutionary conditions and total number of stars there is one and only one value of the integrated luminosity, the only possible source of fluctuation is that related to the total number of stars. This would explain the customary mention to Poisson fluctuations in SBF definitions and its lower relative impact on the EPS literature.

It is now necessary to evaluate the contribution of the Poisson component to the theoretical SBF; only in the case that it provides a measurable contribution, it is justified to mention it in the definition.

To evaluate this contribution, we use $\cal N$ defined by Buzzoni (note that in its original definition $\cal N$ is defined on a Poisson basis, so it uses the second raw moment), and compare this with $\bar{L}_\mathrm{TS88} = \bar{L}_{\Sigma_\mu;N_\mathrm{pois}}$ defined in Eq.  \ref{eq:SBFPoi}:

\begin{equation}
\frac{1}{\cal N} = \frac{\bar{L}_{\mathrm{TS88}}}{\mu_{1}'} = \frac{\mu_{2}'}{\mu_{1}'^2} = \frac{\kappa_{2} + \mu_{1}'^2}{\mu_{1}'^2} = \frac{\kappa_{2}}{\mu_{1}'^2} + 1 = \frac{\bar{L}_{\mathrm{SP}}}{\mu_{1}'} + 1.
\end{equation}

\noindent Hence, the relationship between $\bar{L}_{\mathrm{SP}}$ and $\bar{L}_{\mathrm{TS88}}$
is

\begin{equation}
\frac{\bar{L}_{\mathrm{SP}}}{\bar{L}_{\mathrm{TS88}}} = \frac{\bar{L}_{\mathrm{SP}}}{\bar{L}_{\Sigma_\mu;\mathrm{pois}}} = 1 - {\cal N}.
\label{eq:LtheovsLTS88}
\end{equation}

\noindent $\cal N$ has values around of $10^{-3}$ \citep{Buzz89,Buzz93,CVGLMH02} when the stellar luminosity distribution function is normalized to an initial total mass in the population of 1~M$_{\odot}$, and has a similar order of magnitude when normalized in terms of the number of stars. The numerical values of the pure population SBF, $\bar{L}_{\mathrm{SP}}$, and population SBF plus Poisson contribution due to the variation in the number of stars per resolution element when the reference mean is obtained, $\bar{L}_{\mathrm{TS88}}$, are therefore similar. That is,  {\it Poisson fluctuations in the total number of stars only contribute to around 0.1\% of the total SBF}. As a consequence, it is very ambiguous to make an explicit mention in the definition of SBF a contribution that only accounts for a  0.1\% of the variance, and not mention the intrinsic distributed nature of the integrated light of a stellar population that is responsible for 99.9\% of the SBF. Fortunately for previous SBF studies based on EPS, this change in the definition of SBF has a negligible numerical impact on current SBF computations of Single Stellar Populations models.

This result also shows that (stellar populations) SBF should be an {\it intrinsic result} in EPS studies, and not a particular computation designed for a restricted range of observational cases.

\subsection{Different ways to obtain SBF}
\label{sec:SBFmodes}

Once the nature of the different components implicit in theoretical SBF has been established, it is possible to define different ways to compute SBF. The methods presented here differ mainly in the way the resolution element is defined and how the reference mean is obtained. 

The first is that established by \cite{TS88}, which has been the reference method used in this paper. Its main characteristics are: (a) the resolution elements are the individual pixels in a CCD image; (b)  the reference mean is obtained from a smooth model image of the galaxy profile with a posterior 
treatment of large-scale residuals \cite[we refer to ][and references therein for further details and improvements]{Meietal05}; and (c) the SBF are obtained by means of the spatial power spectrum of the entire galaxy or a galaxy region (in general annular regions around the center). In this method, the main advantages are the use of a large number of resolution elements (the individual pixels), and the use of the power spectrum analysis, which allows a clear discrimination between noise (with a flat power spectrum) and SBF signal.  Additionally, the method includes other sources of fluctuations (such as the contribution of globular clusters) that we have not considered in this work. 
However, the application method is far from simple and there are several  observational aspects that must be taken into consideration. Interested readers may find more detail, for examples,  in  \cite{TS88}, \cite{AT94}, \cite{Tonetal97}, \cite{Jenetal03}, \cite{Meietal05} and references therein. 

A  second method is that proposed by \cite{Buzz07}: (a) The image is divided into several regions following the symmetry of the system (i.e. pie-like sections) and the flux is obtained in each section. These sections (that may contain a large number of pixels) are the resolution elements; (b) The average flux is obtained from all the resolution elements and the renormalized fluctuation with respect to this common mean is computed; (c) The variance of the resulting distribution of renormalized fluctuations is obtained. In this case, the resolution element clearly includes both the SP component and the profile component, although in fact it is just a natural extension of the original method for which, obviously, the Poisson approximation does not apply. Although this method may not provide accurate SBF values, it can be used as a first SBF estimation and for stellar populations studies (provided that the SP component is more important than the profile component).

Other ways to obtain SBF can be proposed, such as the application to resolved systems \citep{AT94,RBCC05}, 
the analysis of the ensemble of IFUs observations of the system as opposed to the analysis of each particular IFU, a combination of both methods, or, even, the use of data archives with different observations of the same system. The crucial points to remember in these situations are: (a) how the stellar population is sampled in each one of the considered resolution elements and (b) how the distribution of the mean number of stars in the resolution elements is used to define a mean reference flux for each element.

\subsection{Stellar population SBF as an intrinsic metric of fit in EPS studies}
\label{sec:SBFuse}

In the previous section we provided different ways to obtain SBF from the data. However, we showed that SBF are an intrinsic result of the modeling of stellar populations that provides information on the dispersion in the integrated luminosity of any stellar system \citep{Buzz93,Buzz07}. Furthermore, stellar population SBF are a scale-free quantity independent of the number of stars in the system (although the shape of the distribution of renormalized fluctuations is not scale free). Hence, stellar population SBF provide an intrinsic {\it metric of fit}  for any comparison of a single observation (e.g. the integrated spectrum of a galaxy) with synthesis models results \cite[see][ for more details]{CL07a,CL07b}.

When a single observation is compared with the {\it mean} obtained by EPS models, it is possible to obtain its associated renormalized fluctuations precisely as achieved in Sect. \ref{sec:populationSBF}, using Eq. \ref{eq:phiSBFteo}. Hence, we can obtain the position of the individual observation in the distribution of renormalized fluctuations $\varphi(L^\mathrm{fluc}_{\mathrm{SP};N^*})$ and establish an associated probability for the corresponding fit. In other words, stellar population SBF provide a tool to establish the {\it accuracy} of the fit, as opposed to current $\chi^2$ methods that only provide {\it precise} values without any accuracy information.

In particular, the comparison of stellar population SBF at different wavelengths infers, that it is  impossible to fit data at infrared wavelengths with the same precision as at optical wavelengths, since the first have a higher SBF value, and hence a larger {\it intrinsic} dispersion. This result is independent of the number of stars in the observation. As a consequence, a probabilistic paradigm in the use of synthesis models is necessary not only for the analysis of systems with a low number of stars (affected by IMF sampling effects) but for an accurate analysis of any system, regardless of size.

\section{Conclusions}
\label{sec:conclusion}

In this work we have investigated the nature of SBF. We have introduced a distinction between stellar population SBF, theoretical SBF, and observational SBF. Stellar population SBF depend only on the intrinsic properties of stellar populations; theoretical SBF are determined by both the population and the contribution from the distribution of the total number of stars in the used resolution element; and the observational SBF represent the observational counterpart of theoretical SBF. Stellar population SBF and theoretical SBF should be considered in probabilistic terms. On the other hand, observed SBF are statistically distributed and should therefore be interpreted in statistical terms.

In this paper, we have applied the probabilistic description of synthesis models to SBF. We have found that SBF are simply a manifestation of the way in which Nature samples either the integrated luminosity distribution function (pLDF) in integrated systems across different resolution elements, or the stellar luminosity distribution function (sLDF) of resolved systems. We have also found that this definition does not require any assumption about the distribution of stars across different stellar types. Stellar population SBF are therefore simply defined to be the ratio  between the {\it variance} and the mean of the luminosity {\it distribution} function in linear fluxes:

\begin{equation}
\bar{L}_{\mathrm{SP}} = \frac{\kappa_{2}}{\mu_{1}'} = \frac{\kappa_{2}(L_{\mathrm{tot};N^*})}{\mu_{1}'(L_{\mathrm{tot};N^*})}.
\end{equation}

In addition to the definition of stellar population SBF, we have introduced a definition for theoretical SBF, which consists in general of a theoretical description of an observational method. We have shown that the effect of the distribution in the number of stars per resolution element, $N$, is included in theoretical SBF in two ways. First, this distribution affects the computation of the mean flux needed to define the amplitude of renormalized fluctuations for each resolution element. Second, the distribution of mean fluxes also affects  theoretical SBF. Both effects introduce a term that is added to the population contribution. Only for flat galaxy profiles can this term be a simple Poisson contribution; in the case of a different profile, the term differs and must be evaluated explicitly. As a result, theoretical SBF are defined to be:

\begin{equation}
\bar{L}_{\mathrm{theo}} = \frac{\kappa_{2}}{\mu_{1}'} + \mu_{1}'\,\frac{\kappa_2(N)}{\mu_{1}'(N)} = \bar{L}_{\mathrm{SP}} + \mu_{1}'\,\frac{\kappa_2(N)}{\mu_{1}'(N)}.
\end{equation}

Observational SBF are an {\it estimation} of theoretical SBF and must be corrected for the contribution of the variation in the total number of stars per pixel, which depends on both the observational strategy (by means of the way in which the local mean is computed) and the particular galaxy (by means of the galaxy profile), before being compared with stellar population SBF.

Assuming, as a particular case, that the total number of stars per resolution element is Poisson-distributed, we have shown that the corresponding contribution is negligible compared with the contribution of  the stellar population in the break-up of theoretical SBF. Therefore, the common theoretical definition of SBF as "mean, luminosity-weighted luminosity of the stellar population'' resulting from a Poisson assumption is a restrictive definition that does not necessary apply to the observed SBF, nor any definitions based on an interpretation of SBF as Poisson fluctuations in the number of stars. 

We have also characterized the distribution of renormalized fluctuations for different cases in terms of skewness and kurtosis, showing their relation with the skewness and kurtosis of the sLDF. We have shown that the shape of the renormalized fluctuation distributions for a system with $N^*$ stars per resolution element is similar to the shape of the integrated luminosity distribution of clusters with $N^*$ stars, the population luminosity distribution function (pLDF). Taking advantage of the results by \cite{CLCL06}, we have shown that the renormalized fluctuations of flux across galaxies are not Gaussian, as had been commonly assumed.

Additionally, we have shown alternative ways of compute SBF and that stellar population SBF define an intrinsic metric of fit for the comparison of single observations with synthesis models results. In fact,  {\it SBF are observationaling evidence that the results of evolutionary synthesis models cannot be interpreted in a deterministic way, but must be interpreted using a probabilistic paradigm.}

Finally, as a caution to theoretical computations of SBF, it is fundamental to ensure that the stellar types considered in the computations are defined in so that the luminosity can be assumed to be constant within each stellar class considered. SBF computed with isochrones that include a detailed treatment of the post-AGB evolution \citep{Cioni06a,Cioni06b} yield only lower limits to real SBF, since the post-AGB sequence provided by these isochrones are the {\it average} luminosity during the post-AGB evolution. The variance in this average luminosity must also be known before realistic SBF computations can be obtained based on these isochrones.

\begin{acknowledgements}
We want to acknowledge useful discussions on observational SBF and photon counting issues by Jaime Perea and Asunci\'on del Olmo. We also thank Gabriella Raimondo and Rosa Amelia Gonz\'alez L\'opezlira for steady discussions on the subject of SBF since 2003 and a thorough reading of the first draft. We also acknowledge Amelia Bayo and the comments from the referee, which have substantially improved the paper.
This work has been supported by the Spanish {\it Programa Nacional de Astronom\'\i a y Astrof\'\i sica} through FEDER funding of the project AYA2004-02703 and AYA2007-64712. MC is supported by a {\it Ram\'on y Cajal} fellowship. LJ acknowledges support by a contract of the AYA2004-02703 project.
\end{acknowledgements}

\bibliographystyle{apj}

\end{document}